\documentclass[fleqn,twoside]{article}
\usepackage{espcrc2}

\usepackage{epsfig}

\newcommand{\AmS}{{\protect\the\textfont2
  A\kern-.1667em\lower.5ex\hbox{M}\kern-.125emS}}

\title{QCD in Extreme Environments.}

\author{J.~B.~Kogut\address{Physics Department, University of Illinois,
                            1110 West Green Street, Urbana, IL 61801, USA}
\thanks{
JBK is supported in small part by an NSF grant NSF PHY-0102409.
}}

\begin{document}
\begin{abstract}
I review present challenges that QCD in extreme environments presents to
lattice gauge theory. Recent data and impressions from RHIC
are emphasized. Physical pictures of heavy ion wavefunctions, collisions and
the generation of the Quark Gluon Plasma are discussed, with an eye toward engaging
the lattice and its numerical methods in more interaction with the experimental and
phenomenological developments. Controversial, but stimulating scenarios which can be
confirmed or dismissed by lattice methods are covered. In the second half of the talk,
several promising developments presented at the conference Lattice 2002 are reviewed.
\end{abstract}

\maketitle

\section{Introduction}

We live in interesting times.

In the world of high energy and nuclear physics the Relativistic Heavy Ion
Collider (RHIC) at Brookhaven National Laboratory is beginning its search
into the new realm of high temperatures and low but nonzero chemical potentials.

These experiments will surprise us. Experiments have a habit of doing that. They
humble us. They will show us new directions. They will make this talk obsolete.

I want to emphasize that Lattice Gauge Theory, which has become an evolutionary
rather than a revolutionary field, should take a more active part in these developments. 
It should rise to the challenge of real data to find new methods and ideas and extend
its comfortable Euclidean ways to describe, predict and learn from scattering
experiments. Pioneering has the potential of reinvigorating the field. Lattice gauge
theory has provided a solid estimate for the critical temperature to make the 
quark gluon plasma as well as not-so-solid estimates of the required energy density,
the magnitudes of
screening lengths etc. But there is much more to predict...hopefully before the
experiments...and doing so will be particularly exciting. And fun.

I think that there are promising areas for such developments and I will discuss
some throughout this talk. Light-cone wavefunctions of nuclei, dense with gluons, quarks
and anti-quarks, collisions with unexpectedly high multiplicities and signs of
early development of a quark-gluon plasma will be reviewed. Along the way suggestions
for additional or new lattice based research will be made. Other subfields of
high energy physics are already interacting with the data. Lattice gauge theorists
should become more active participants.

It hardly needs to be emphasized here that when Lattice Gauge Theory develops a
method to analyze a problem, it can do so from first principles, with ultimately
no approximations. Emphasis on the word "ultimately", because our lattices are yet
too small, our quark masses are yet too large, our statistics are yet too few, etc.
But the field is making steady progress on all these issues, helped particularly
by its use of improved but practical Actions. Lattice gauge theory is the only
approach that can handle the physics of the three crucial length scales of hadronic
dynamics at one time, in a consistent calculation. At sufficiently short distances
lattice calculations and simulations confirm asymptotic freedom, which is so essential
to exposing the underlying gluon and quark degrees of freedom of QCD as experiments
enter the quark-gluon plasma phase. At more moderate distances where the running coupling
is in the intermediate range and semi-classical instanton configurations are breaking
the anomalous $U(1)_A$ symmetry and, through properties of their ensemble, are breaking
chiral symmetry and are producing the constituent quark masses, lattice gauge theory
is at its best elucidating the complex crossover physics of this range of length scales. Finally,
at greater length scales, Lattice methods confirm confinement, the fact that hadronic states
are color singlets and the confining dynamics comes through thin, discernable but breakable flux tubes.
QCD will not be fully understood until these three qualitatively different ranges
of phenomena are incorporated into one tractable analytic approach. Crucial hints needed
to accomplish this will come from lattice studies.

These are grand, and over-stated words. But progress is occurring.

It is a pity that progress is not occurring on the challenge of producing a lattice simulation
method for QCD at nonzero Baryon chemical potential $\mu$ and vanishing temperature $T$. 
The infamous sign problem of
the fermion determinant continues to stand in our way. The same problem has bedeviled condensed
matter physicists for almost 40 years. 

A theme of this talk is that studying extreme environments
teaches us how QCD works under ordinary conditions. I believe that when we have licked the sign
problem, or have found a new formulation of nonperturbative QCD free of it, we will be at a new
level in our understanding of how QCD really works and makes its low lying Baryonic excitations.

In the second half of this review I will discuss several interesting contributions at this
conference. These will include progress in mapping out the low $\mu$ (chemical potential),
high $T$ (temperature) part of the phase diagram of QCD, following the seminal work of
Fodor and Katz. Real time spectral functions for the production of lepton pairs will
be discussed as will the dispersion relation of pions below but near the transition
to the plasma. A first step toward simulating phenomenologically interesting cutoff Four
Fermi models of the transition will also be mentioned briefly.

\section{What's so Great about the Quark Gluon Plasma?}

As emphasized by E. Shuryak \cite{shuryak}, the mass scales of the Quark Gluon Plasma are different and, importantly,
smaller than those of the more familiar hadronic phase. The hadronic phase breaks chiral symmetry, the
quark gluon plasma does not. The hadronic phase confines quarks, the quark gluon plasma does not. The
binding mechanism in the hadronic phase is nonperturbative while the screening mechanism in the 
quark gluon plasma is perturbative.

We know from deep inelastic scattering that the substructure scale in the hadronic phase is $Q^2 \approx
1-2$ GeV. At this $Q^2$ the running coupling is becoming large enough that perturbation theory is failing
to give accurate estimates and nonperturbative effects are competitive.

By contrast perturbative screening masses in the quark gluon plasma are $M^{gluon} \approx 0.4$ GeV. and
$M^{quark} \approx 0.3$ GeV. for temperature above but near $T_c$ where the plasma first appears.

The finer level spacings in the plasma act as a fine resolution grid to the dynamics in the hadronic phase.
A collision which starts in the hadronic phase and ends in the plasma phase is especially sensitive to
the levels and dynamics of the hadronic phase because of the relative wealth of open channels in the plasma phase.

\section{Overview of RHIC Collisions}

RHIC collisions can be analyzed in four stages which are thought to be characterized by distinct
time scales.

First, viewed from the center of momentum frame, there are right moving (positive $p_z$)
and left moving (negative $p_z$) nuclei. Each of
them can be described by infinite momentum frame wavefunctions which give the probability amplitudes
for each nuclei to consist of ensembles of quarks and gluons. The parton model, suitably improved 
with asymptotic freedom to describe its short distance features, should provide a framework for
these initial states.

Second, there is the collision between the constituents in each nuclei. Experiments will shed light on
the typical momentum transfers and multiplicities of these underlying collisions.

Third, there is the development of a thermalized plasma of quarks and gluons.

And fourth, there is the development of the final states of hadronic debris from the hot soup of colliding
and produced constituents.

\subsection{The Initial State}

Begin with the lightcone wavefunctions of the projectiles. The rightmover is,

\begin{eqnarray}
|\Psi>=\sum_n \int \prod_i^n \frac{d^2 K_i d \eta_i}{\eta_i} 
\psi_n(\eta_1,\vec{K}_1;...;\eta_n,\vec{K}_n) \nonumber \\
\delta^2(\sum \vec{K}_i) \delta(1-\sum \eta_i) |\eta_1,\vec{K}_1;...;\eta_n,\vec{K}_n>
\end{eqnarray}

Recall from infinite momentum frame \cite{ks} or light cone perturbation theory \cite{ks} that
$\eta_i$ is the longitudinal fraction of the i$^{th}$ constituent. It is a positive variable,
$0 < \eta_i < 1$ and the constituents in each wavefunction $\psi_n$ account for the
longitudinal fraction of the projectile, $\sum_i^n \eta_i = 1$. The absence of vacuum structure
in $|\Psi>$ makes this wavefunction a useful formulation of the initial state of the collision.

The bulk features of the wavefunctions of the projectiles are conveniently displayed on a rapidity plot,
$r=\frac{1}{2} \ln [(p_0+p_z)/(p_0-p_z)]$. The rapidity is particularly handy because it simply
translates under a boost along the $z$ axis. One can plot the density of partons of both projectiles
on a rapidity plot and deep inelastic experiments at a fixed $Q^2$ indicate that the plot is 
essentially flat and boost invariant and of extent $\ln s$, where $s$ is the usual scattering variable,
four times the square of the center of momentum energy.

\begin{figure}[htb!]
\centerline{
\epsfxsize 3 in
\epsfysize 3 in
\epsfbox{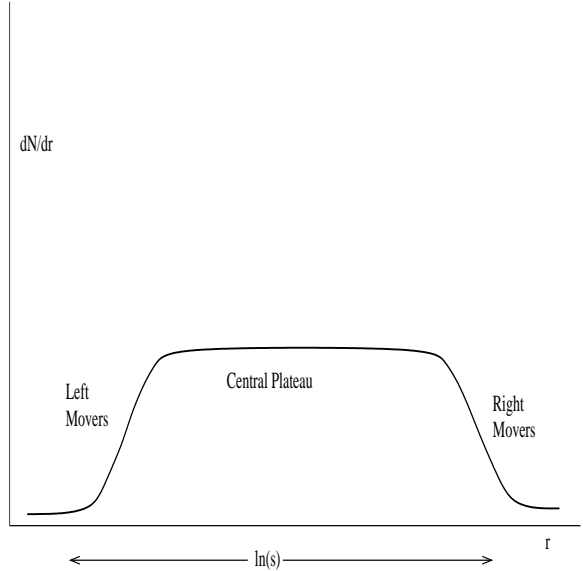}
}
\caption{Rapidity distribution of partons in a nucleus-nucleus collision.
}
\label{fig:rapidity}
\end{figure}

Experiments indicate that parton rapidity distributions have some simple, statistical features.

1. Short range correlations in rapidity. The parton distributions lose memory, with a
correlation length of $1-2$ units, in rapidity. Screening in rapidity is an essential feature.
Therefore, features like the quantum numbers of
each projectile are limited to the edges of the rapidity plot. Unfortunately, this means that in
very high energy heavy ion collisions, most of the partons are not influenced by the Baryon number of
the projectiles. The collisions can be characterized by a high energy density and temperature but
not high chemical potential $\mu$.

2. The "central region" and is universal and "vacuum like". The length of the central region grows
like $\ln s$.

Although first principle calculations of light cone wavefunctions are beyond us, models having these features
have been known for decades. Multiperipheral graphs of QCD can populate the rapidity axis uniformly
and have suggested that gluons are the most likely parton species in the central region \cite{lipatov}.

Asymptotic freedom effects these considerations profoundly. Recall that in the naive parton
model, partons interact only softly with one another, never exchanging high transverse momenta. But
if we consider the structure of the wavefunctions at high spatial resolution where the running
coupling is small and perturbative asymptotic freedom applies, then high transverse momentum exchanges and
short distance additional structure are inevitable \cite{ks2,ap}. Call the gluon distribution function in momentum space
$\eta G(\eta,Q^2)$. This is the distribution function that would be resolved by a probe with
resolving power $\delta X_{\perp}^2 \sim \frac{1}{Q^2}$. Models and theoretical analysis of
scattering data suggests that $\eta G(\eta,Q^2)$ approaches a constant for small $\eta$ and fixed $Q^2$.
However, as $Q^2$ is increased one discovers that gluons can split perturbatively into three or four
gluons of lesser longitudinal fractions $\eta_i$. So, for sufficiently small $\eta$, $\eta G(\eta,Q^2)$
must be an increasing function of $Q^2$. This is a general feature of the asymptotically free parton
model \cite{ks2} and perturbative evolution equations make this simple physical idea quantitative \cite{ap}. For quark
distribution functions which are accessible to  deep inelastic electron and neutrino scattering, these
ideas are well tested phenomenologically.

If we view the parton density on the plane transverse to the collision axis $p_z$, then at low $Q^2$ we see
just a few gluons of size $\delta X_{\perp}^2 \sim \frac{1}{Q^2}$, but as $Q^2$ is taken larger we 
resolve many more gluons of smaller size. This progressive development has lead to the idea of "gluon
saturation" in the light cone wavefunction of a nucleus \cite{iancu}.

To understand this idea, consider the wave function and imagine that it is 
coming at you and you have a snapshot of the transverse plane.
Call the transverse size of the nucleus $R_A$. Each parton of transverse momentum $Q$ in the wavefunction
occupies an area $\pi/Q^2$, as suggested by the uncertainty principle. The parton's cross section is proportional
to the square of the running coupling and its geometrical size, $\sigma \sim \alpha_s(Q^2) \frac{\pi}{Q^2}$.
But partons will overlap in the transverse plane when their number $N_A$ becomes comparable to 
$\frac{\pi R_A^2}{\sigma}$ which is $\frac{Q^2 R_A^2}{\alpha_s(Q^2)}$. One supposes that when $N_A$ is this large,
the overlap
stops further growth in the density and "saturation" has been achieved \cite{iancu}. The transverse momentum of the
partons at saturation is $Q_s^2 \sim \alpha_s(Q^2) \frac{N_A}{R_A^2} \sim A^{1/3}$. The number of
partons at the saturation level is $N_s \sim \frac{Q_s^2 R_A^2}{\alpha_s(Q_s^2)} \sim A$ and these are
the number of partons of size scale $1/Q_s$ which could materialize in the quark gluon plasma. In addition $Q_s^2$
sets the scale for the transverse momenta of partons in the central region of the rapidity plot,
$\frac{1}{\sigma} \frac{dN}{dr d^2 K_{\perp}}=\frac{1}{\alpha_s(Q_s^2)} F(\frac{K_{\perp}}{Q_s})$.

These formulas can be made more precise by considering longitudinal fractions \cite{iancu}. The discussion here is just
a simplified, but, hopefully, intriguing introduction. When numbers for the various nuclei and the energies
available at experimental facilities are substituted here, one finds that $Q_s^2 \sim 1-2$ GeV. This means that
applying perturbative QCD is suspect and competitive nonperturbative effects should be expected. This
is potentially the realm of lattice gauge theory. The relationship of Euclidean lattice calculations and the
Minkowski scattering formulation appropriate to heavy ion collisions must be addressed.

\section{Space Time Evolution of the Collision}

Let's comment on the four different stages of a heavy ion collision separately. The physics issues
in each are quite distinct.

1. Wave functions approach on another.

As reviewed above, the phase space parton density is quite large $\sim \frac{1}{\alpha_s(Q_s^2)}$. Classical and
semi-classical estimates should apply. In particular, the color per volume is large, so the lack of
commutivity, $[Q^a,Q^b]=if^{abc}Q^c$, appears to be a relatively small effect, and the gluon field can be treated
classically for many purposes \cite{iancu}.

2. Interactions.

The gluon saturation picture suggests that the characteristic time of interactions is quite small,
$t_{int} \sim \frac{1}{Q_s} \sim 0.1-0.3$ fm./c and the energy density is very, very large,
$\epsilon_{int} \sim \frac{Q_s^4}{\alpha_s(Q_s^2)} \sim 20$ GeV./fm.$^3$. Such a large energy density
would place the system deep in the quark gluon plasma phase. This is good news! Estimates of $\epsilon_{int}$
in the past were much more modest, on the order of several GeV./fm.$^3$, at most.

3. Thermalization.

After the collision the quark gluon matter expands and thermalizes. The experimental data and models suggest
that the time scale for this to occur is $t_{therm} \sim 0.5-1.0$ fm/c.

RHIC data, especially that showing the spatial distribution of particles produced in non-head-on collisions,
strongly suggests that the interactions at early times are strong and the quark gluon plasma appears very
early in the evolution of the final state \cite{kharzeev}. The basic parton-parton interactions must be strong and
involve high multiplicities. Perturbative processes alone are not sufficient to explain the data.

4. Hydrodynamic expansion.

The quark gluon plasma maintains thermalization and expands until decoupling sets in at $t_{dec} \sim 10$fm./c.
In the expansion and production of the final state hadrons, the fastest particles are produced last, as
in the "inside-outside" cascade of electron-positron annihilation processes \cite{bj}. The expansion process is
essentially one-dimensional, along the collision axis, until the latest stages of the development of the
final state. A particle's rapidity turns out to be strongly correlated with the space-time rapidity,
$\frac{1}{2} \ln [(x_0+x_3)/(x_0-x_3)]$, of its point of creation.

\section{Suggestions from RHIC and Evidence for a Quark Gluon Plasma}

RHIC experiments are hard and complicated and subject to many phenomenological models. Typically several
models, each in stark contradiction to the other, can describe a limited set of heavy ion data equally
well. The "smoking gun" for the existence of the quark gluon plasma has yet to be found, but there are several
pieces of data which are nicely explained, or at least, accommodated by it. Consider a few pieces of the puzzle.

1. Jet Quenching.

The heavy ion collisions show little or no sign of jets for $p_{\perp} < 4$ GeV. This is interesting
in light of the fact that if you scale single particle hadron spectra from proton-proton data to A-A data,
you overshoot the jet data very significantly \cite{kharzeev}.

It has been argued that this is evidence for the existence of an extended space-time region of the
quark gluon plasma. The idea is that  two colliding gluons in the plasma annihilate into an energetic
quark anti-quark pair with sizeable tranverse momenta, but their interactions with the plasma medium saps
the energy out of the energetic quark and its partner, eliminating ("quenching") those jet-like characteristics.
Of course, very high energy jets are uneffected by  the medium because of the short range nature of
the most prevalent interactions on the rapidity axis, and jets with energies much larger than $4$ GeV.
are seen.

2. Suggestions of Deconfinement and Chiral Symmetry Restoration.

Lepton pairs have long been cited as effective probes into the internal dynamics of the quark gluon
plasma. In particular, if the $\rho$ and $\phi$ existed in the plasma, then the leptonic decays,
$\rho \rightarrow ee$ and $\phi \rightarrow ee$ would be easily seen, as in proton-A collisions.
These peaks are missing, however \cite{kharzeev}. It is tempting to interpret this experimental result as evidence
that the light hadrons have "melted" in the hot plasma.

We will consider some lattice gauge theory data concerning these peaks in the second half of this talk.

3. Screening and J/$\psi$ Suppression.

One of the first theoretical suggestions for a signal of the existence of the quark gluon plasma concerned
the heavy quark states J/$\psi$. It was argued and backed up with potential model calculations, that
quasi-free quark and gluon thermal screening would reduce the attractive forces between the heavy quarks 
sufficiently to eliminate the binding energies and the J/$\psi$ states themselves \cite{matsui}. 
In fact, the suppression
of these states, again implied by the absence of the associated lepton pairs, is much more dramatic than
that expected in hadron absorption models.

\section{Is there Nonperturbative, Semi-Classical Physics at Work in Heavy Ion Collisions?}

RHIC experiments suggest that there are early, strong, high multiplicity parton-parton collisions 
in heavy ion scattering processes. Perturbation theory, such as the QCD multiperipheral diagrams \cite{lipatov} which
can fill the rapidity plot with a uniform density of gluons, fall far short of describing the data. This
suggests that larger length scales where the running coupling is in the intermediate coupling range
are involved. This is the domain where semi-classical collective excitations of QCD are important.
E. Shuryak has suggested that instantons and sphalerons \cite{shuryak2} may be causing some of the surprising scattering events
seen at RHIC.

\section{What is a Lattice Theorist to do?}

We have taken a short survey into topics of interest to heavy ion physicists. Can lattice theorists have
much impact here? Much of the physics here is stated in Minkowski space, natural to scattering
processes. Employing Euclidean Lattice methods to tackle some of these questions will certainly
take new initiatives and methods. As we will review in the second half of this talk, lattice methods do
provide information on real time processes even within the present techniques we use. These include
power spectra, Maximum Entropy Methods and lepton pair rates. However, more
comprehensive methods closer to first principles can probably be developed, too.

Let's look at a few such topics.

\subsection{Lattice Approach to Nucleus Wave Functions}

The parton picture of QCD structure requires the use of the infinite momentum frame, or, equivalently
light cone quantization. Each parton has a particular tranverse and longitudinal momentum in a wavefunction
expansion, as written above. Infinite momentum frame Hamiltonians have been written down in this language and
two dimensional Hamiltonians which control the transverse dynamics have been studied. Lattice gauge theory versions
of these systems can be constructed \cite{krasnitz} and equivalent three dimensional Lagrangian systems derived.

One can hope that formulations of this sort could be used to address issues of parton saturation in the context of
lattice methods which can handle honestly the intermediate coupling character of these problems. The
infinite momentum frame idea will eliminate vacuum fluctuations from the nonzero longitudinal fraction parts
of the problem.

In the context of continuum methods, there have been recent attempts along these lines by heavy ion physicists.
An ambitious attempt to formulate deep inelastic scattering in Euclidean terms has been pioneered
by O. Nachtmann \cite{nachtmann}.
This looks like particularly fertile ground for a lattice incursion.

\subsection{Nonperturbative Physics for $T \geq T_c$ and small $\mu$}

As I will review in the second half of this talk, the region of the QCD phase diagram for
small $\mu$ in the vicinity of the quark gluon transition $T_c$ is accessible to conventional simulation methods.
This is a recent development that opens up experimentally relevant lattice calculations to the whole field.
The thermodynamics of the formation of the quark gluon plasma for $T$ near $T_c$ and small Baryon
chemical potential $\mu$ should be studied very professionally now. Some of the projects will
be reviewed in the second half of this talk.

We would certainly like to know how quickly the underlying physical mechanisms responsible for the
transition from the hadronic phase to the quark gluon plasma phase change
as $\mu$ is turned on for $T \approx T_c$. We know that there are substantial
interactions in the quark gluon plasma just above $T_c$. How are these mechanisms and how are bulk thermodynamic
quantities effected as $\mu$ is taken nonzero?

The role of instantons in the formation and character of the quark gluon plasma needs more elucidation
even at the $\mu=0$ edge of the QCD phase diagram.
According to the instanton liquid model \cite{schaefer}, the instanton ensemble is disordered in the hadronic phase. This is
essential for the model to produce chiral symmetry breaking and phenomenologically reasonable
numbers. It is believed
that as $T$ passes through $T_c$, instantons and anti-instantons pair up and form "neutral" molecules. Chiral
symmetry is restored in such an ordered ensemble.

It would be interesting to verify this "binding-unbinding" picture of the phase transition by lattice methods
and to provide estimates of the scales of the phenomena involved. This will not be easy because
instantons and anti-instanton molecules will tend to "fall through" the lattice and be hard to identify, let alone
measure. Improved actions will probably be needed.

Such studies would be worth doing because we know from other simulations that there are substantial
interactions in the quark gluon plasma just above $T_c$. We would also like to know how instanton dynamics
change as $\mu$ is turned on for $T \approx T_c$.

These are interesting questions which are experimentally relevant. Conventional lattice methods
can address them now.

\subsection{Diquarks and Color Superconductivity?}

Perturbative \cite{Wilczek} and instanton \cite{shuryak4} methods have been used to predict color superconductivity for
low temperatures and asymptotically large $\mu$. A large gap has been found as well as interesting
symmetries ("color-flavor locking" \cite{Wilczek}). The large gap in the spectrum of the Cooper quark-quark pairs
suggests that this state of matter should occur in natural phenomena, such as the interiors of neutron stars,
at high but not inaccessible densities. The lattice has been unable to contribute to this field
because of the sign problem in the Euclidean fermion determinant in $SU(3)$ lattice gauge theory.

The weak coupling, perturbative and instanton, approaches to color superconductivity have not addressed
the numerical impact of confinement on their considerations, so estimates of a critical density
are suspect. Much of the research has concentrated on $SU(2)$ color which is qualitatively different
from QCD. Cooper pairs in the $SU(2)$ model are color singlets and are degenerate with
ordinary meson states because of the self-conjugate nature of quark and anti-quarks in this case. At chemical
potentials greater than half the pion mass, the ground state of the model becomes a superfluid.

Instanton enthusiasts have noted that their model provides attractive forces which favor the
appearance of scalar $\bar{3}$ diquarks \cite{rapp}. The forces are only half as strong as the analogous
ones in the $SU(2)$ model where diquark states are degenerate with mesons. This suggests one
should find considerable scalar $\bar{3}$ diquark correlations in ordinary hadronic physics. To my knowledge,
there are none. Perhaps lattice spectroscopists could scour their baryon wavefunctions and search for diquark
correlations in their three body states.

Aside from the issue of color superconductivity, it would
be good to settle the old controversy concerning the possible viability of diquark models
of baryon structure. 

Weak coupling descriptions of color superconductivity don't need diquark correlation enhancements in
ordinary hadronic states to be successful and accurate at nonzero chemical potentials. This is because the
Cooper pairs of the BCS theory have very large, "floppy" wavefunctions which are much larger 
than protons \cite{rajagopal}.

\section{Contributions to Lattice 2002}

The parallel sessions at Lattice 2002 concerning QCD in extreme environments contained a number of 
interesting contributions. Because of the lack of space I will discuss just four topics that generated
the most discussion. I will have to point the reader to the writeups of the parallel talks for details.

The four topics will be : 1. The high $T$, low $\mu$ edge of the QCD phase diagram, 2. Thermal dilepton rates,
3. Real-time pion propagation in hot QCD, and 4. Four dimensional Four Fermi models at nonzero $\mu$.

\subsection{High $T$, low $\mu$ edge of the QCD phase diagram}

Following the seminal work of Fodor and Katz last year \cite{fodor}, the field has realized that the
high $T$, low $\mu$ edge of the QCD phase diagram is accessible to a variety of mature lattice
techniques. Fodor and collaborators \cite{fodor2} presented arguments that their multi-parameter generalization
of the Glasgow method has an efficiency which only falls off as a power,
$V^{-\alpha}$ with $\alpha \approx 0.2-0.3$, as the size of the lattice grows large. 
de Forcrand and Philipsen \cite{philipsen}
have done simulations at imaginary chemical potentials using conventional Hybrid Molecular Dynamics
algorithms and have found convincing, controlled results for $\mu_q < \pi T_c/3$. 
The Bielefeld-Swansea group \cite{allton}
have presented Taylor expansions in $\mu$ of the Fodor-Katz method. D.K. Sinclair \cite{dks} simulated $SU(3)$ lattice
gauge theory at nonzero isospin chemical potential $\mu_I$ and found that $T_c(\mu_I)$ resembles
the curve $T_c(\mu)$ predicted by the other groups.

Before we consider the first two contributions in more detail, recall some elementary points. The transition
from the hadronic phase to the quark gluon plasma phase at $T=T_c$ and $\mu=0.0$ is actually a
rapid crossover for the range of quark masses presently under investigation. Most lattice theorists
believe that it is also a rapid crossover for realistic quark masses, $m_u$, $m_d$ and $m_s$, but larger scale
simulations are needed here. As $\mu$ is turned on one expects that the crossover $T_c(\mu)$ to
decrease slowly. However, for sufficiently large $\mu$ one expects the crossover to become a first
order transition and it is this point "E", the critical endpoint, $T_E$ and $\mu_E$, that one would 
like to know \cite{fodor}.
Present simulations indicate that $\mu_E$ is quite large \cite{fodor}. However, those simulations are on small
lattices with relatively large quark masses. They should be repeated on large lattices with realistic
quark masses to get a relevant answer for $T_E$ and $\mu_E$. If $\mu_E$ turns out to be very small,
there would be dramatic effects, potentially, in the projectile fragmentation regions at RHIC.

\begin{figure}[htb!]
\centerline{
\epsfxsize 3 in
\epsfysize 3 in
\epsfbox{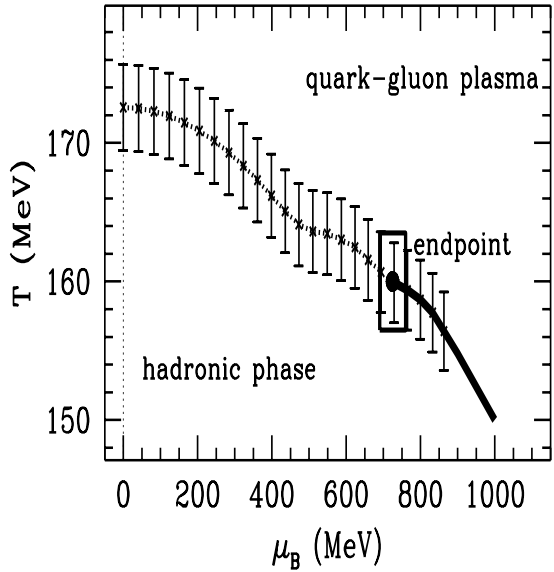}
}
\caption{Evolution of the crossover to the endpoint of a first order line of transitions as $\mu_B$ increases.
}
\label{fig:endpoint}
\end{figure}

Fodor and Katz \cite{fodor} introduced multi-parameter reweighting to improve the overlap of lattice configurations
on the edge of the phase diagram with particular points inside the phase diagram,

\begin{eqnarray}
Z(m,\mu,\beta)=\int DU e^{-S_{bos}(\beta_0,U)} \det M(m_0,\mu=0,U) \nonumber \\
\{ e^{-S_{bos}(\beta,U)+S_{bos}(\beta_0,U)} \frac{\det M(m,\mu,U)}{\det M(m_0,\mu=0,U)} \}
\end{eqnarray}

This formula organizes the calculation of the Partition function at a point inside the phase
diagram, by making configurations on the edge of the phase diagram (first line) and reweighting (second line)
in two variables, $\mu$ and $\beta$. The idea is that configurations at $T_c$ experience both the
hadron and the quark gluon plasma phase so there should be a considerable overlap with configurations
nearby but at different $\mu$ and $\beta$ values. Care must be taken here : as $\mu$ is increased, $\beta$
must be increased slightly
so the system remains near the crossover coupling for that $\mu$ value,
$\beta_c(\mu)$. As $\mu$ and $\beta$ are increased, one monitors the acceptance rate and
keeps it near 50 percent \cite{fodor2}. In this way lines essentially parallel to $\beta_c(\mu)$ but displaced
into the hadronic or the plasma phase have also been investigated and the equation of state has been calculated
for a range of $T$ and $\mu$ values. They have shown that the crossover sharpens as $\mu$ grows and
the endpoint "E" is approached \cite{fodor2}. The position of the endpoint was found in last year's contribution
but studying the finite size scaling properties of the Lee-Yang zeros of the Partition function \cite{fodor}.

\begin{figure}[htb!]
\centerline{
\epsfxsize 3 in
\epsfysize 3 in
\epsfbox{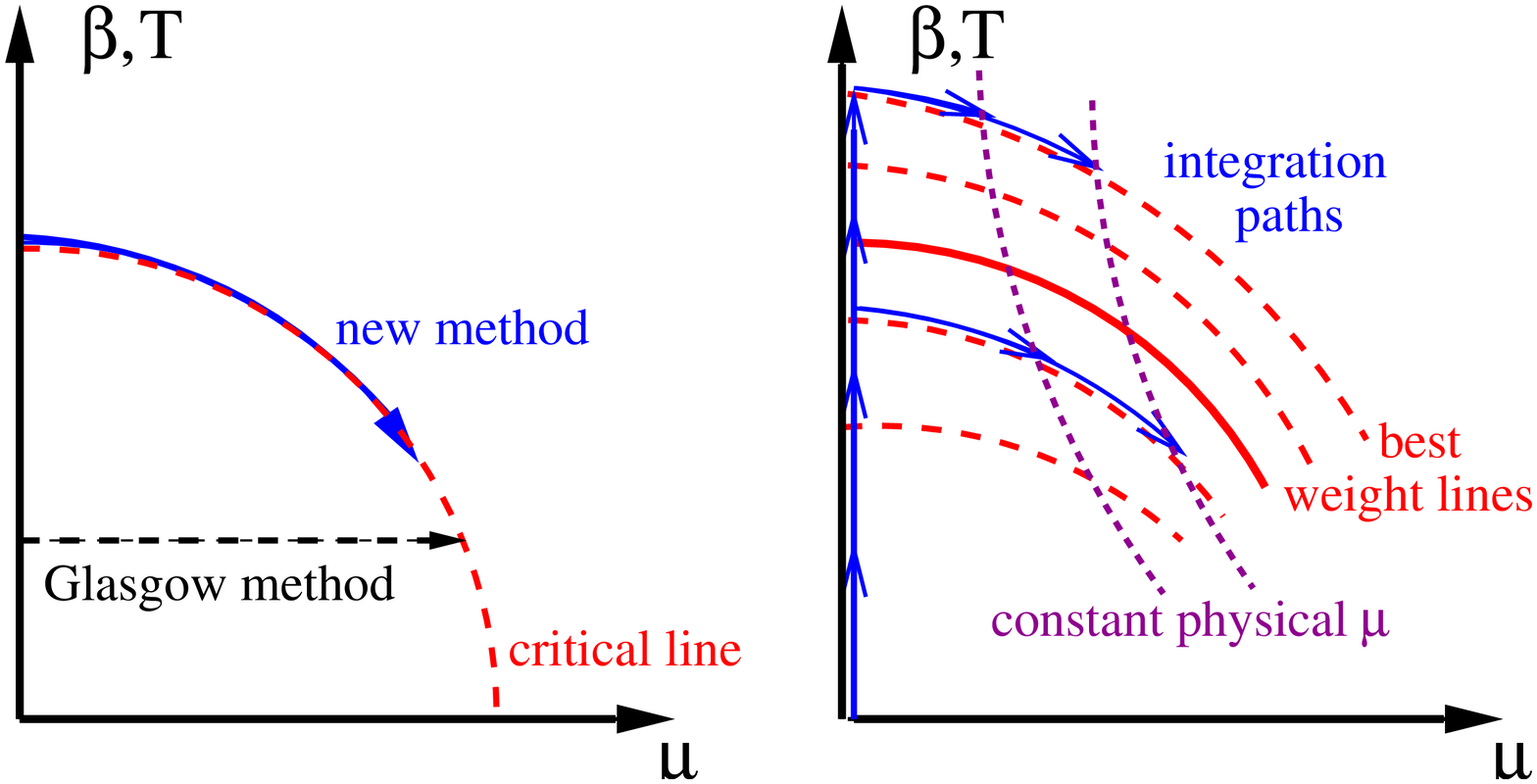}
}
\caption{Paths in the phase diagram used by Fodor and Katz compared to Glasgow's original method.
}
\label{fig:path}
\end{figure}

Following the paths in phase space where the authors monitor the efficiency of their overlap method, they are
now computing the Equation of State \cite{fodor2}. The crossover is
predicted to sharpen considerably as $\mu_B$ increases.

The authors argue that at least for volumes up to $12^3 \times 4$, they can simulate chemical potentials
up to $\mu \propto V^{\gamma}$ with $\gamma \approx 0.2-0.3$. And they argue that the efficiency of their
multi-parameter overlap method scales as $V^{-\alpha}$ with $\alpha \approx 0.2-0.3$, so the method
should allow fairly large lattices to be simulated before its inefficiencies become overwhelming.

The reader should consult the parallel session contribution by these authors for plots and details \cite{fodor2}.

The second contribution presented on this topic by de Forcrand and Philipsen \cite{philipsen}
simulated lattice at imaginary chemical potential using the
conventional hybrid molecular dynamics algorithm. It followed the crossover, 
$\beta_c(\mu)$, by computing the plaquette susceptibility and monitoring its peak as a function
of $\beta$ for fixed values of the imaginary chemical potetial.

The authors argue that the curve $\beta_c(\mu)$ is an even, analytic function of $\mu$. Writing 
$\beta_c(\mu)=\sum_n c_n (a\mu)^{2n}$, they find that only the first two terms are needed
to describe the data accurately. The analytic continuation to real $\mu$ is, therefore, trivial.
Their resulting curve is in agreement with that of Fodor and Katz, although they haven't investigated
the order of the transition as a function of $\mu$ yet.

The imaginary chemical potential method is very conservative and effective. It will work on large lattices
with the efficiency of an ordinary hybrid molecular dynamics code. It can only treat small chemical potentials
because imaginary chemical potentials induce tunnelling to other $Z(3)$ vacua when the imaginary chemical potential
equals $\pi T/3$ \cite{weiss}. For $T$ near $T_c \approx 170$ MeV., the method is limited to $\mu \leq 500$ MeV.

The reader is referred to the relevant contribution at the parallel session for quantitative details. In addition,
similar work has appeared by D'Elia and Lombardo \cite{lombardo}.

We can look forward to considerable progress in understanding this strip of the QCD phase diagram now that
several simulation methods and groups are involved.

\subsection{Thermal dilepton rates}

Dilepton rates were discussed above as an informative probe into the dynamics of the quark gluon plasma.
This topic also provides a nice example of traditional lattice gauge theory confronting new experimental
scattering data.

The dilepton rate, which is a function of frequency $\omega$ and momentum $\vec{p}$, can be written 
in the form \cite{karsch2},

\begin{eqnarray}
\frac{dW}{d \omega d \vec{p}}=\frac{5 \alpha^2}{27 \pi^2}
\frac{1}{\omega^2 (e^{\omega/T}-1)} \sigma_V(\omega,\vec{p};T)
\end{eqnarray}

\noindent where the spectral function $\sigma_V(\omega,\vec{p};T)$ is related to the current-current
correlation function, $G_V(\tau,\vec{p};T)=\int d \vec{x} e^{i \vec{p} \cdot \vec{x}} 
<J_V^{\mu}(\tau,\vec{x})J_{V,\mu}^{\dagger}(0,\vec{0})>$, on the periodic lattice as,

\begin{eqnarray}
G_V(\tau,\vec{p};T)=\int d \omega \sigma_V(\omega,\vec{p};T)
\frac{\cosh (\omega (\tau-1/2T)}{\sinh (\omega/2T)}
\end{eqnarray}

On a $N^3 \times N_{\tau}$ lattice, one can measure the correlation function $G_V$ at each "time" slice,
$\tau_k T= k/N_{\tau}$, $k=1,2,...,N_{\tau}$ and then attempt to invert the relation between the 
spectral function and the correlation function.

There are two approaches to executing this inversion. First, one might use a model dependent functional form
which depends on several parameters and determine $\sigma_V(\omega,\vec{p};T)$. This approach 
has had only limited success. Alternatively, one might use the Maximum Entropy Method (MEM) \cite{asakawa}
and find the
"most probable" spectral function. Since the lattice simulation gives the correlation function only at a discrete set
of points $\tau_k$, each with statistical errors, the problem of finding the continuous function
$\sigma_V$ of the continuous variable $\omega$ is ill-posed. The Maximum Entropy Method can only produce a
"most probable" answer with a measure of its reliability. The method, however, is well-known and quite
successful in optics, where it is used for image reconstruction. Again, the reader should consult 
the parallel sessions for a wealth
of examples that illustrate the method's strengths and weaknesses.

At past conferences the Bielefeld group has presented data \cite{karsch2} which have shown that the $\rho$ and $\phi$
peaks in the dilepton spectra are suppressed in the presence of the quark gluon plasma, in qualitative 
accord with the experimental trends. However, the dilepton signal is still somewhat larger than that found for
free but hot quarks and gluons, at least for $\omega/T$ between 4 and 8. These simulations were done above
the transition, at $1.5T_c$ and $3 T_c$, so measureable attraction in these light quark states at these
high temperatures is somewhat surprising. Results such as these suggest that although bulk thermodynamic behavior
of the quark gluon plasma might be close to the Stephan-Boltzmann limit, detailed hadronic structures
are not completely washed out.

At this conference S. Datta \cite{datta} presented Bielefeld simulation results at $0.9 T_c$ and $1.2 T_c$ for the
charmonium states. The S-wave states had a diminished, perhaps by a factor of 2, signal above the transition
where potential models predicted there would be no signal at all. The P-wave signal, however, was reduced
by a factor of about 7 across the transition suggesting that screening has removed these charmonium states
from the spectrum.

\subsection{Real-time pion propagation in hot QCD}

A study of Effective Lagrangians by Stephanov and Son \cite{son} has uncovered curious features in the propagation
of pions as the quark gluon plasma is approached by heating hadronic matter, $T \leq T_c$.
In fact, the pion dispersion relation is found to be determined by temperature dependent features of the system
which are accessible to conventional Euclidean lattice simulations. The three features one needs are 
1. the pion screening mass, 2. the pion decay constant, and 3. the axial isospin susceptibility.

This work constitutes a nice, informative example where "real-time" quantities are accessible to static
simulations. Of course, the most popular example of this is the relation of the speed of sound $v_s$ in a system
to its bulk thermodynamics, $v_s^2=\partial P/ \partial \epsilon$, where $P$ is the bulk pressure and $\epsilon$
is the energy density. Other examples exist in condensed matter physics, as well. For example, the
propagation features of spin waves in anti-ferromagnets are determined in large part by static correlation functions.

These authors \cite{son} determine the real part of the soft pion dispersion relation,

\begin{eqnarray}
\omega_p^2=u^2 (\vec{p}^2+m^2)
\end{eqnarray}

\noindent where we identify $u$, the pion's speed, $m$, the pion's screening mass, and $\omega_{p=0}=um$,
the pion's "pole mass". The authors derive three low energy theorems at nonzero $T$. First, $u^2=f^2/\chi_{I5}$, where
$f$ is the pion decay constant and $\chi_{I5}$ is the axial isospin susceptibility. These last two 
quantities are related by the other two theorems, which represent the generalizations of the Gell-Mann Oakes Renner 
relation to finite temperatures, $f^2m^2=\chi_{I5} m_p^2=-m_q<\bar{\psi} \psi>$. Since these quantities
are accessible to conventional lattice simulations, we have a nice illustration of determining "real time"
physical quantities from static, Euclidean simulations.

The strategy of this development is interesting and quite general. One begins with the microscopic
quark Lagrangian underlying lattice QCD,

\begin{eqnarray}
L_q=i\bar{\psi} \not{D} \psi -m_q \bar{\psi} \psi+\mu_{I5} \bar{\psi} \gamma_0 \gamma_5 \frac{\tau_3}{2} \psi
\end{eqnarray}

\noindent as well as the low energy effective Lagrangian that describes the properties of the low energy mesons,

\begin{eqnarray}
L_{eff}=\frac{f_t^2}{4} Tr \nabla_0 \Sigma \nabla_0 \Sigma^{\dagger}-
        \frac{f_s^2}{4} Tr \partial_i \Sigma \partial_i \Sigma^{\dagger} \nonumber \\
	+\frac{1}{2} m^2f_s^2 Re Tr \Sigma
\end{eqnarray}

\noindent where $\Sigma=\exp (i\tau^a \pi^a/F_{\pi})$, and 
$\nabla_0 \Sigma = \partial_0 \Sigma-\frac{i}{2} \mu_{I5} (\tau_3 \Sigma + \Sigma \tau_3)$. The Effective
Lagrangian should be familiar from strong interaction phenomenology and its extension to nonzero
chemical potential is determined by the symmetries of the underlying microscopic Lagrangian without any
numerical or phenomenological ambiguity \cite{Toublan}.

By matching the predictions of the microscopic theory with those of the low energy Lagrangian, the authors
determine their low energy theorems.

Note how curious pion propagation is as $T$ approaches $T_c$. The pion propagation speed appraoches zero, as does
the pion decay constant $f$. A further investigation of the scaling laws governing the approach to the quark gluon
transition reveals that $u$ vanishes faster as $T$ approaches $T_c$ than the screening mass $m$ diverges \cite{son},
so the pole mass $m_p=u m$ approaches zero as $T$ approaches $T_c$. Since the pion's speed vanishes in
this limit while its pole mass does also, perhaps we can call the pion an "infinitely massive massless" state!
Hopefully, some of these features will show up in final state studies at RHIC.

\subsection{Four dimensional Four Fermi models at nonzero $\mu$}

As advertised in the first part of this talk, it would be good for all concerned if there were more
interaction between the lattice community and the
phenomenologists interested in the quark gluon plasma and in color superconductivity. One step in this direction has
been taken by the Swansea group \cite{walters} which has found diquark condensation in strongly cutoff $3+1$ dimensional
Nambu Jona-Lasinio models, of the sort discussed by Nuclear theorists, instanton phenomenologists and others.
The value in the lattice simulations is, as usual, that their results are exact in principle, 
because they include all the fluctuations.

The Lagrangian which has been simulated is just one of the original Nambu Jona-Lasinio models,

\begin{eqnarray}
L_{NJL}=\bar{\psi}_i(\not{\partial}+m_0+\mu \gamma_0) \psi_i -\frac{g^2}{2}[(\bar{\psi}_i\psi_i)^2- \nonumber \\
(\bar{\psi}_i \gamma_5 \vec{\tau} \psi_i)^2]+
\frac{i}{2}[j(\bar{\psi}_i^{tr} C \gamma_5 \tau_2 \epsilon_{ij} \psi_j)+ \nonumber \\
\bar{j}(\bar{\psi}_i C \gamma_5 \tau_2 \epsilon_{ij} \bar{\psi}_j^{tr})]
\end{eqnarray}

\noindent where the last two terms are source terms for diquark condensation which are taken to zero 
($j \rightarrow 0$) at the end of a simulation meant to calculate the diquark condensate. These terms play the
same role as a small bare fermion mass term in a lattice simulation of the chiral condensate---the simulation must
pick out a preferred direction in the space of the condensate, calculate the condensate and see if it survives
in the limit where the source is taken away. In fact, simulations of this model indicate that as $\mu$ increases,
a diquark condensate appears in the system and its $\mu$ dependence and magnitude are close to the predictions
of the leading term in the large $N$ expansion. The reader should consult the parallel sessions of these conference
proceedings for plots \cite{walters}.

One notable feature of this calculation is that it is the first lattice simulation
in $3+1$ dimensions with a traditional fermi surface at nonzero chemical potential. Future work in
this direction will be the study of additional four fermi models suggested by the instanton liquid model.
One complication will be the cutoff nature of these models---these models are logarithmically trivial
and they are useful only when strongly cutoff, with their parameters adjusted so that observables
such as the pion decay constant and the chiral condensate take on physical values. Luckily, the Nuclear
physics community has used such models for some time and know the "best" way to deal with these issues.
The lattice model may reach an impasse, however, because when complicated four fermi interactions are
entertained, it becomes impossible to avoid complex fermion determinants. In addition, these models
do not confine quarks in their low temperature phase, so they are subject to criticisms discussed earlier.
Nonetheless, these studies should be informative and represent a test of some of the analytic,
approximate approaches to the high $\mu$, low $T$ superconductivity phase.

\section{Conclusions}

Lattice gauge theory has had much to say about the finite temperature behavior of QCD because it
can handle the three length scales of QCD, perturbative behavior at weak coupling and short distances,
semi-classical coherent behavior at intermediate couplings and intermediate distances, and confinement
at strong couplings and large distances. It is now able to study the variable $T$-low $\mu$ strip of the phase
diagram and make predictions relevant to heavy ion experiments. I hope that its lessons will lead to additional
formulations and strategies to study real time physics relevant to collisions and nuclear structure
in the not-too-distant future.

\end{document}